\documentclass[preprint,authoryear,12pt]{elsarticle}
  
\usepackage{amssymb}

\journal{Arxiv}

\begin{document}

\begin{frontmatter}


\tnotetext[label1]{}
\author{Anil Raghav\corref{cor1}\fnref{label2}}
\ead{raghavanil1984@gmail.com}
\author{Ankush Bhaskar\fnref{label3}}
\author{Virendra Yadav\fnref{label3}}
\author{Nitinkumar Bijewar\fnref{label2}}
\cortext[cor1]{corresponding author}
\address{1. Department of Physics, University of Mumbai,Vidyanagari, Santacruz (E), Mumbai-400098, India.\fnref{label2}}
\address{2. Indian Institute of Geomagnetism, Kalamboli Highway, New Panvel, Navi Mumbai- 410218, India.\fnref{label3}}

\title{Low energy secondary cosmic ray flux (gamma rays) monitoring and its constrains}
 





\begin{abstract}

Temporal variation of secondary cosmic rays (SCR) flux was measured during the several full and new moon and days close to them at Department of Physics, University of Mumbai, Mumbai (Geomagnetic latitude: 10.6$^\circ$ N), India. The measurements were done by using NaI (Tl) scintillation detector with energy threshold of 200 keV. The SCR flux shows sudden enhancement for approximately about 2 hour in counts during couple of events out of all experimental observations. The maximum Enhancement SCR flux is about $ 200 \%$ as compared to the diurnal trend of SCR temporal variations. Weather parameters (temperature and relative humidity) were continuously monitored during all observation. The influences of geomagnetic field, interplanetary parameters and tidal effect on SCR flux have been considered. Summed spectra corresponding to enhancement duration indicates appearance of atmospheric radioactivity which shows single gamma ray line. Detail investigation revealed the presence of radioactive $Ar^{41}$. This measurements puts limitations on low energy SCR flux monitoring. This paper will help many researcher who want to measure SCR flux during eclipses and motivate to find unknown mechanism behind decrease/ increase during solar/lunar eclipse.     

\end{abstract}


\end{frontmatter}


\section{Introduction}

In recent times, SCR flux (mainly low energy gamma ray flux) variations during solar and lunar eclipses have attracted attention. A typical decrease in SCR flux during solar eclipse and increase in SCR flux during lunar eclipse has been reported by \cite{Ananda67, Bhattacharyya97, Kandemir00, Antonova07, Bhattacharya10, Nayak10, Bhaskar11} and \cite{Raghav13}. To understand the underlying physical mechanism behind the decrease of SCR flux during solar eclipse, researchers have attempted to correlate local weather parameters and geomagnetic variation with SCR flux. For example, \cite{Bhattacharyya97} have ascribed the decrease in SCR flux during a solar eclipse to atmospheric cooling. However, \cite{Chintal97} have ascribed the observed decrease in $\gamma $ ray flux to blocking of the Sun by the Moon. \cite{Bhaskar11} have ascribed the gradual variations in SCR flux by scintillation detector to the blocking of the Sun and observed time delayed response in GM counter measurements, attributing it to physical processes occurring in the atmosphere during the eclipse.

\cite{Ananda67} had studied lunar and solar eclipses by monitoring variation in SCR flux using a  Geiger-Muller (GM) counter. He had observed increase in SCR flux during the lunar eclipses \cite{Ananda67}. Confirmation of this had been done by \cite{Raghav13} using scintillation detector and they have ruled out the possibility of local weather, interplanetary and geomagnetic parameters, which at first appeared to be the likely candidates causing the enhancement in SCR flux. The unique geometrical alignment of the Sun, the Earth and the Moon during an eclipse and effective gravitational force causing tides in magnetospheric plasma and crustal deformations may be responsible for the observed variations in SCR flux during eclipses. Researchers have reported observations which are explained based on tidal theory. \cite{Krymsky62}  has reported that the SCR variation with a period of half a lunar day could be explained by the lunar tidal effect based on diurnal SCR variations from neutron and muon monitors. \cite{Dorman61} have noticed that during Full/New Moon SCR flux is enhanced/decreased. They have interpreted this as a possible lunar tidal effect on the Earths magnetospheric plasma which cuases variations in geomagnetic rigidity. However, the cuase of SCR variations is not yet well established. 

\cite{Volodichev91} have reported atmospheric radioactivity (an intensity burst of thermal neutrons) during a solar eclipse. They have also shown that thermal neutron enhancement occurs during New and Full Moon and the days close to them. They have ascribed the enhancement to the crossing of lunar tidal wave over their observation site. This causes deformation of cracks in the Earth's crust which releases trapped radioactive gases, mainly Radon, into the atmosphere. The alpha particles generated by Radon gase interact with the crust and the atmospheric atoms/molecules, giving rise to the observed neutron splashes\cite{Volodichev87,Volodichev91,Volodichev97,Antonova07}.

To investigate the underlying physical mechanism behind decrease during solar eclipse and enhancement during lunar eclipse, its important to study temporal variation of SCR flux during New and Full Moon and the days close to them. Therefore, we have carried out measurements of SCR flux at Department of Physics, University of Mumbai, Mumbai (Geomagnetic latitude: 10.6$^\circ$ N), India. Measurements were done using a NaI (Tl) scintillation detector having dimensions of 7.62 cm $\times$ 7.62 cm. The details of the observational setup are described in \cite{Raghav13}. Ambient temperature and relative humidity were recorded at every 5 minute interval during the observations using a digital temperature and humidity sensor. 


%

\section{Observations and Interpretations}

\label{}
\begin{figure}[htp]
\centering
\includegraphics[angle=0,width=140mm]{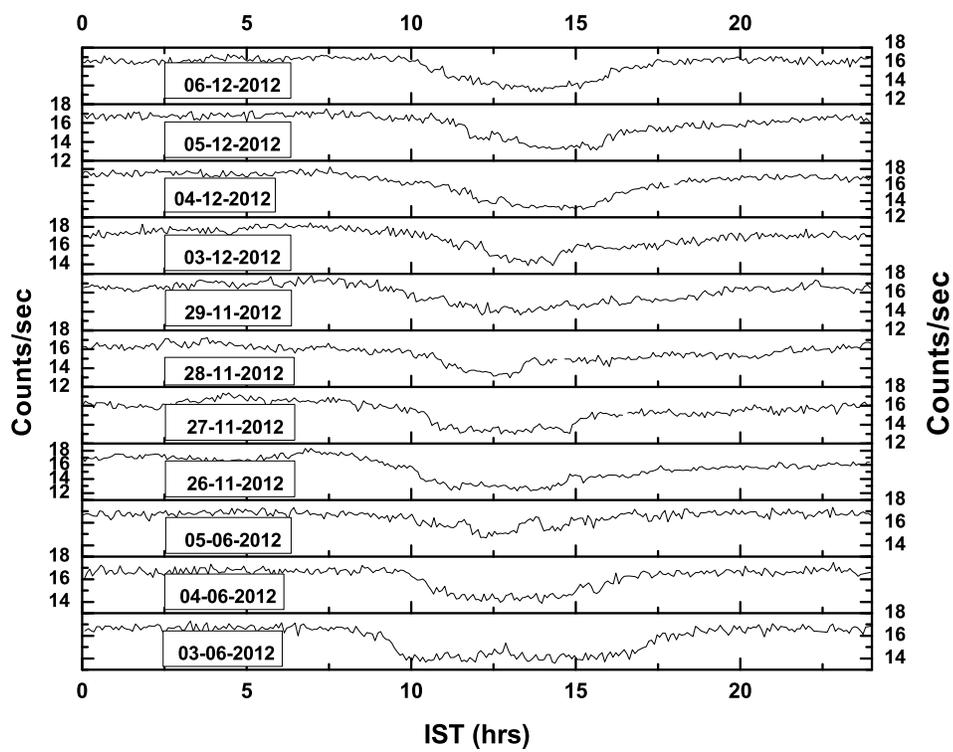}
\caption{Diurnal variation of SCR flux on control days close to full and new moon.}
\label{fig:diurnal} 
\end{figure}

The regular diurnal variations of SCR flux were measured for 45 days during 2012-2013, close to the full and new moon as per the weather conditions. To demonstrate the diurnal profile of SCR flux, temporal variations are shown for few selected days in Figure ~\ref {fig:diurnal}. Note that, in general, profile of diurnal variation shows minimum during afternoon where as maximum and steady flux is observed during the night. Diurnal variations show that decrease starts after the sunrise which accompanies increasing/decreasing trends in temperature/humidity. Similarly, increasing trend starts before the sunset which accompanies decreasing/increasing trends in temperature/humidity. This is consistent with the negative/positive correlation of SCR flux with temperature/humidity as reported by \cite{Raghav13}. 

\begin{figure}[htp]
\centering
\includegraphics[angle=0,width=140mm]{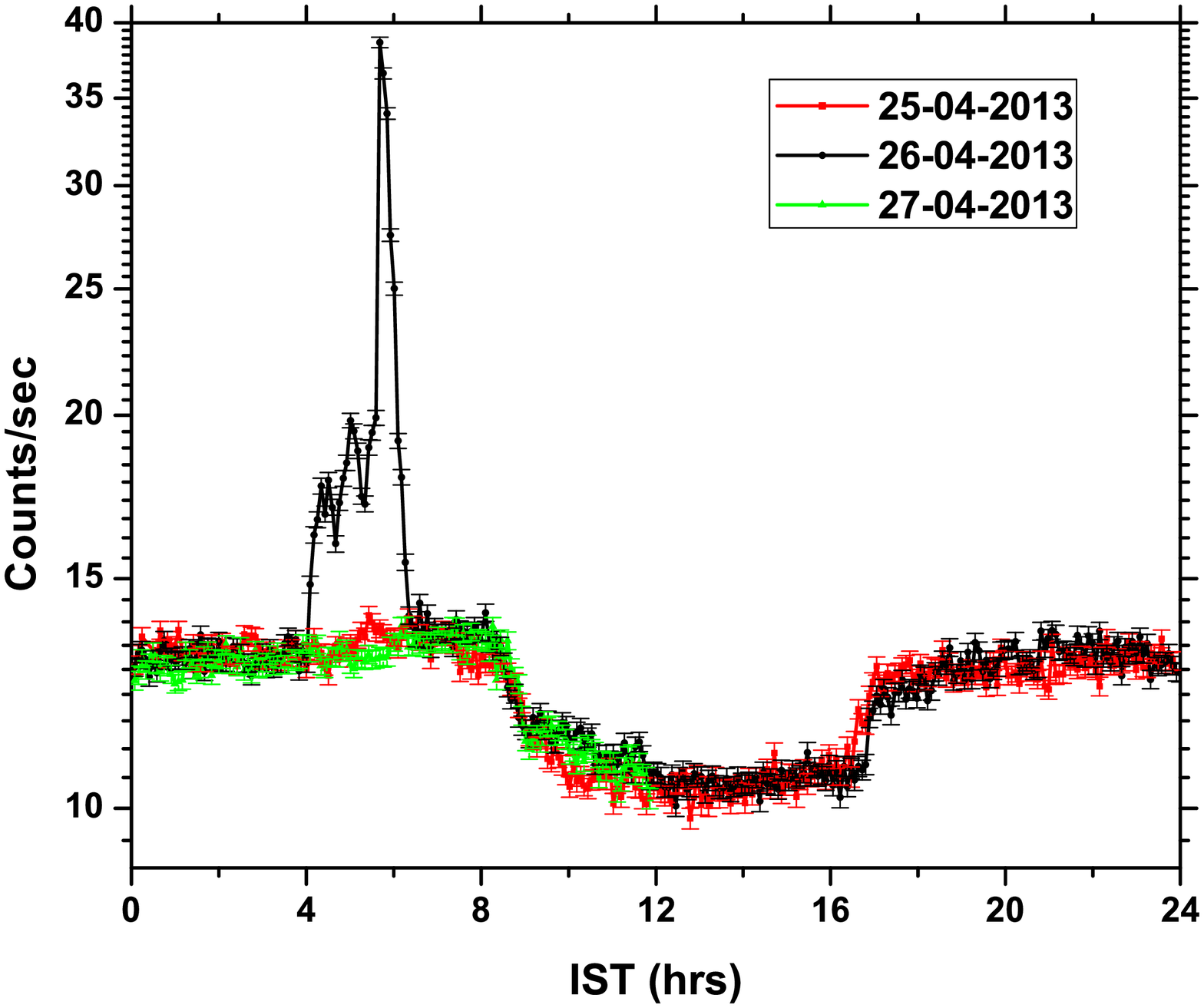}
\caption{Abnormal enhancement on April 26, 2013 as compare to adjacent control days.}
\label{fig:26april} 
\end{figure}

\begin{figure}[htp]
\centering
\includegraphics[angle=0,width=140mm]{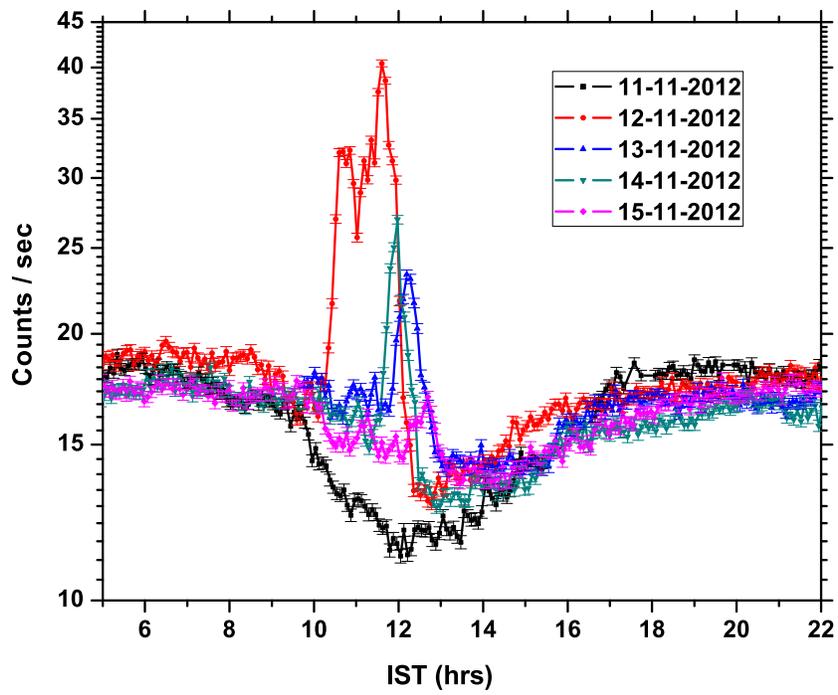}
\caption{Abnormal enhancement on November 12-15, 2013 as compare to a control day.}
\label{fig:12nov} 
\end{figure}

There were two abnormal events on days close to the full and new moon which show sudden enhancement in SCR flux for approximately two hours. The Figure ~\ref{fig:26april} and ~\ref{fig:12nov} show these abnormal events. On April 26, 2013 we observed flux increase in early morning around 04:00 to 06:00 (local time) but there is no signature of the enhancement on previous or following days of the event. The maximum increase is approximately $200 \%$  as compared to the regular steady flux observed just before/after the event and the SCR flux at the corresponding time of the previous and following days. Whereas, the other event showed enhancement consecutively for four days during November 12-15, 2012 which occurred close to local noon on each day. The amplitude and duration of enhancement shows day to day variation.
After observing Figure ~\ref{fig:26april} and ~\ref{fig:12nov}, one may ascribe this enhancement purely to noise and completely overlook the observations. Such events might have been observed in the past but their interpretation as noise would have constrained the communication of the same. At first, even authors and some researchers with whom authors discussed also thought the enhancement was due to noise. Normally, detector electronics or pick up from connecting cables give rise to high shoot-up in the counts of lower energy region of the spectrum. We have analyzed all raw spectrum data files and confirmed that there is no shoot up in the low energy counts during the these abnormal events. 


As we have confirmed that this is not at all noise, its important to search the source of this sudden enhancement in SCR flux. Atmospheric (temperature and humidity), geomagnetic (Kp index and SYM-H index ) and interplanetary (interplanetary magnetic field and solar wind pressure) parameters were studied during both the events. Also we have checked  MOSCOW neutron monitor data. However, no abrupt changes were seen in any of the above mentioned parameters during any of the events (not shown here).

\begin{figure}[htp]
\centering
\includegraphics[angle=0,width=140mm]{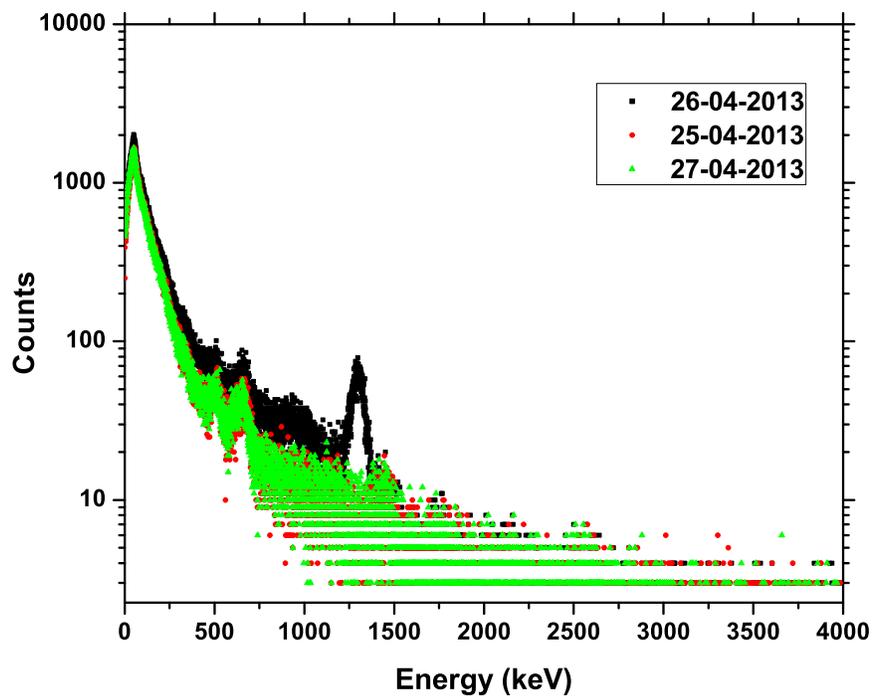}
\caption{Summed spectra for April 26, 2013 event duration (4:00 to 6:00 IST) and corresponding duration on control days. Black dots represent event day whereas red and green dots represent control days}
\label{fig:26add-spectra} 
\end{figure} 

\begin{figure}[htp]
\centering
\includegraphics[angle=0,width=140mm]{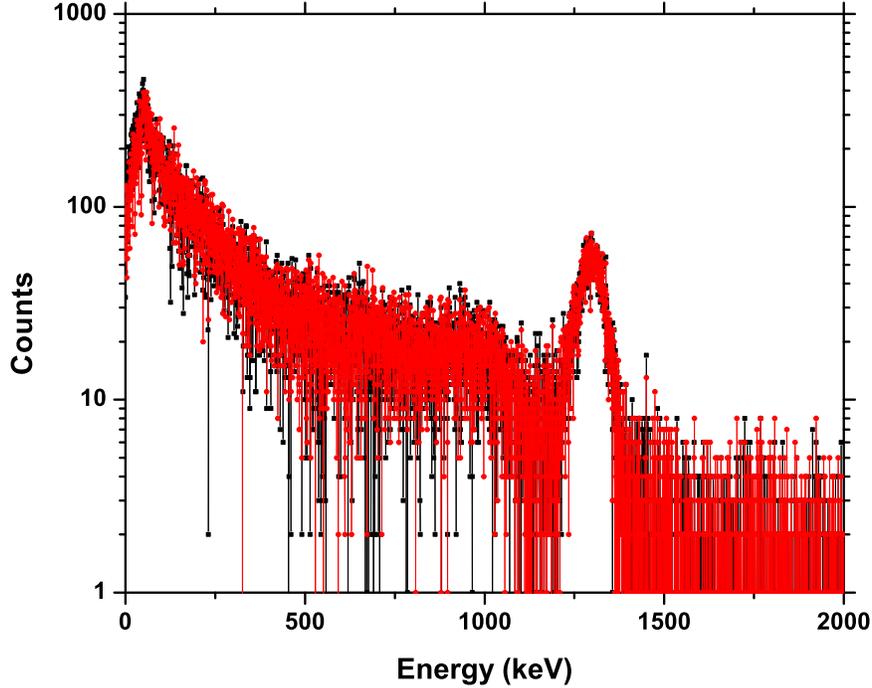}
\caption{subtraction of the summed spectra of previous and following days from the summed spectrum of event day. Red represents event day $\-$ previous day and black represents event day $\-$ following day }
\label{fig:26sub-spectra} 
\end{figure}

To analyzed the energy spectrum during both the events, all the spectrum files corresponding to the each event duration were added. The summed spectra for April 26, 2013 and adjacent control days are shown in Figure ~\ref{fig:26add-spectra}. In the figure, an enhanced peak is observed which might be the main contributing factor in the abnormal enhancement in SCR flux. To confirm this, the summed spectra of previous and following days were subtracted from the summed spectrum of event day. The subtracted spectra shown in Figure ~\ref{fig:26sub-spectra} indicate presence of a single gamma ray line with its photo-peak and corresponding compton continuum.
The energy of the gamma ray line was estimated to be $1.29 \pm 0.03$ MeV) by calibration of the detector setup. Similar type of analysis was done for the other event. All of them show presence of the same gamma ray line but having different amplitudes on different days. 

%
%

To identify the source element of the gamma ray line different databases of nuclear decay schemes were scanned. We narrowed down to the nuclei which emit the gamma ray of energy ranging from 1.25 to 1.35 MeV. While identifying potential candidate of source we put the following condition on the sources: source should emit only single gamma ray line or the transition probability of one gamma ray emission in the given energy window should be maximum and much larger than the other transition probabilities. $Na^{22}$ is a strong candidate as a possible source since it emits single gamma ray of energy 1.274 MeV. However, the corresponding essential annihilation peak (511 keV) due to positive beta emission from $Na^{22}$ is not observed in the subtracted spectra. Thus $Na^{22}$ is not the source for the observed enhancement. Another strong candidate is $Ar^{41}$ which emits a single gamma ray of energy 1.293 MeV. The observed energy peak is 1.29 MeV in subtracted spectra which is very close to the $Ar^{41}$ gamma line. Other than these we could not find any other potential candidate. Therefore, we conclude that the source of enhancement of atmospheric radioactivity during the event is $Ar^{41}$.

\section{Discussion and conclusion}

After confirming the energy of single gamma ray emission line we investigated the possible generation mechanism of $Ar^{41}$. In the atmosphere non radioactive $Ar^{40}$ is present in free state. It is possible to convert $Ar^{40}$ to $Ar^{41}$ by neutron activation in which thermal neutron is captured by the nucleus of the $Ar^{40}$. There are two sources from which thermal neutrons can be generated and eventually may lead  to the production of $Ar^{41}$, (1) by natural neutron bursts or (2) anthropogenic sources.  

\cite{Volodichev91} reported the presence of thermal neutron burst during full and new moon and the days close to them. These neutrons may interact with $Ar^{40}$ and produce $Ar^{41}$. But they have postulated that the emission of Radon from the Earth's crust which interact with the atmospheric molecules and produce neutron burst. However in the present study we have not observed any other gamma ray line other than $1.29 MeV)$. That means no signatures of radon gas or its byproducts seen in the observed spectra. So even though observation site is situated in the region having 23 fault lines, the mechanism predicted by \cite{Volodichev91} appears not the possible mechanism. Also, the possibility of release of radioactive gases from the sea cannot be neglected in which $Ar^{41}$ can be one of the radioactive gases. When, the cloud/plume of such gases pass over the site then detector observes the characteristic emission line. But still this is unlikely to be a source as one should observe other gamma ray lines from accompanying radioactive gases.

The second most probable source for $Ar^{41}$ is nuclear reactor. The site of observation is approximately 10 km away from India's two research reactors, Cirus (40MW) and Dhruva (100 MW). In nuclear reactors $Ar^{40}$ is used as a coolant. Emitted thermal neutrons from the reactor interact with the coolant $Ar^{40}$ and converts it into the $Ar^{41}$ by capturing neutron. $Ar^{41}$ has half-life time of 110 minutes. The plume of $Ar^{41}$ gas is released in the atmosphere when reactor is in the operation state. The plume dynamics and the transportation is mainly governed by the atmospheric wind flow direction and dynamics at the released altitude \cite{Chatterjee13}. So the transportation of $Ar^{41}$ to the site depends on wind flow direction and dynamics on the observation days and local time. Note that not on all full/new moon days we observed gamma ray line of $Ar^{41}$. This might be due to the unfavorable wind conditions which inhibit the transportation of the $Ar^{41}$ plume to the observation site. So the occurrence rate of $Ar^{41}$ emission line observed at the site mainly depends on the operation status of the detector and wind conditions. The occurrence on new/full moon might be coincidence and may not have direct relationship with proposed physical mechanism by\cite{Volodichev91}. Still one needs to investigate this characteristic in future with detail study.

Since 1995, many low energy gamma ray (SCR) observations during solar eclipse have been reported. Generally control days were chosen as previous or following day of eclipse. The long term monitoring of temporal variation of SCR suggests that there are many parameters which affect temporal variation of SCR such as local weather parameters (temperature, relative humidity and pressure), geomagnetic and interplanetary parameters.  Gamma ray emission from radioactive Radon gas and $Ar^{41}$ are important noise sources for low energy gamma ray monitoring during eclipses. If these sources are strong one can easily identify the contamination by studying the spectrum of SCR  but if contribution due to these sources is small then it is almost impossible to identify them. 

\cite{Raghav13} reported the enhancement in SCR flux during lunar eclipse. In that work they could remove the contribution from radon gas by putting the threshold condition of beta energy emitted by neutron decay. We have revisited the spectra obtained during the lunar eclipse to check the possible contamination due to $Ar^{41}$ but we have not observed any signature of $Ar^{41}$. However $Ar^{41}$ contamination can not be completely rule out as the $Ar^{41}$ source might be very weak to see clear signature of corresponding single gamma ray emission line. If we put threshold of 1.3 MeV then the count statistics of the data become very poor to infer anything conclusively. So to summarize  $Ar^{41}$ adds constrains to low energy SCR monitoring. Earlier reported SCR observations during solar/lunar eclipses might have contamination from $Ar^{41}$ if the observations are made in the region which has nuclear reactor/reactors. In future, measurements of SCR flux during eclipses should be done in nuclear reactor free region or with energy threshold condition of $>1.3 MeV$ when observations are done near nuclear reactors. This puts constrains on choosing observation site and the accessible energy window.

It is important to note that we have eliminated the possibility of Volodichev's thermal neutron burst effect and correlated nuclear reactor effusion of $Ar^{41}$ as the most probable reason behind the 200 $\%$ enhancement of SCR. However, one should not neglect the possibility of unknown atmospheric radioactivity phenomenon with $Ar^{41}$ as its  signature other than nuclear reactor origin. Detail investigation is required to understand these rarely observed atmospheric radioactivity events.

\bibliographystyle{elsarticle-harv}
\bibliography{<your-bib-database>}







\end{document}